\documentclass[fleqn,twoside]{article}
\usepackage{espcrc2} 
 
\usepackage{graphicx} 
\usepackage{epsfig} 

\usepackage[figuresright]{rotating} 

\usepackage{amsmath}
\usepackage{amssymb}

\newcommand{\der}{\ensuremath{{\operatorname{d}}}}
\newcommand{\Ord}{\ensuremath{{\cal O}}}

\newcommand{\gev}{\ensuremath{\mathrm{GeV}}}
\newcommand{\gevsq}{\ensuremath{\mathrm{GeV^2}}}
\newcommand{\braket}[2]{\ensuremath{\left\langle{#1}|{#2}\right\rangle}}
\newcommand{\bra}[1]{\ensuremath{\left\langle{#1}| \right.}}
\newcommand{\ket}[1]{\ensuremath{\left. |{#1}\right\rangle}}
\newcommand{\pif}{\ensuremath{(\pi^+,\pi^0,\pi^-)_f}}
\newcommand{\pii}{\ensuremath{(\pi^+,\pi^0,\pi^-)_i}}
\newcommand{\qf}{\ensuremath{\vec{q}_f}}
\newcommand{\qi}{\ensuremath{\vec{q}_i}}
\newcommand{\rf}{\ensuremath{\vec{r}_f}}
\newcommand{\ri}{\ensuremath{\vec{r}_i}}
\newcommand{\sigtot}{\ensuremath{\sigma_{\rm tot}}}
\newcommand{\sigcex}{\ensuremath{\sigma_{\rm cex}}}
\newcommand{\sigabs}{\ensuremath{\sigma_{\rm abs}}}
\newcommand{\Ptot}{\ensuremath{P_{\rm tot}}}
\newcommand{\Pcex}{\ensuremath{P_{\rm cex}}}
\newcommand{\Pabs}{\ensuremath{P_{\rm abs}}}

\newcommand{\abs}{\ensuremath{a}}
\newcommand{\rin}[1]{\ensuremath{\rho_{\rm in}^{(#1)}}}
\newcommand{\rout}[1]{\ensuremath{\rho_{\rm out}^{(#1)}}}
\newcommand{\Nout}[1]{\ensuremath{N_{\rm out}^{(#1)}}}
\newcommand{\Nin}[1]{\ensuremath{N_{\rm in}^{(#1)}}}
\newcommand{\psiout}{\ensuremath{\psi_{\rm out}}}
\newcommand{\fbar}{\ensuremath{\bar f}}
\title{Pion absorption and rescattering in the ANP model revisited} 
 
\author{I. Schienbein\address{Theoret. Physik II, Univ. Hamburg,  %\\  
                Luruper Chaussee 149, 22761 Hamburg, Germany} and % 
        J.-Y. Yu\address{Theoret. Physik III, Univ. Dortmund, 44221 
Dortmund, Germany}}
\begin{document} 
 
\begin{abstract} 
Single pion leptoproduction in the region of the $(3,3)$ resonance
is currently of high interest for at least two reasons:
(i) These reactions constitute an important part of the total
cross section in low energy reactions and are utilized to detect neutrino
oscillations in current and future long baseline experiments.
(ii) Intranuclear rescattering of the pions in heavy nuclei
results in interesting and sizable modifications of the free
nucleon cross sections which are testable in electroproduction
experiments.
In this article we give a basic introduction to the pion multiple
scattering model of Adler, Nussinov, and Paschos (ANP) with
special emphasis on pion absorption. We also estimate the probability
of multiple scattering.
\vspace{1pc} 
\end{abstract} 
 
\maketitle 
 
\section{Introduction} 
\label{sec:intro}
Neutrino nucleus scattering will be an important
tool in future long baseline (LBL) experiments 
to precisely measure neutrino parameters like 
neutrino masses and mixing angles.
These experiments will use neutrino beams with
energies of the order $\Ord(1\ \gev)$ where
quasi-elastic reactions (QE), single pion resonance
production (RES) and deep inelastic scattering (DIS)
are all important.
Moreover, due to the small neutrino cross sections
heavy targets like oxygen, argon or iron have to be used.

In the following we will deal with single pion resonance 
production, i.e., with the reactions 
\begin{equation}
\nu +T\rightarrow l +T^\prime +\pi^{\pm,0}
\label{eq:reaction}
\end{equation}
where $T$ is a nuclear target ($_8O^{16},\, _{18} Ar^{40},\, _{26} Fe^{56}$)
and $T^\prime$ a final nuclear state.

% Factorization assumption
% Figure 2.14 in Ji-Young's thesis
At this stage we make the basic assumption that the
reaction 
\eqref{eq:reaction} can be described by 
two {\em independent} steps \cite{Adler:1974qu}:
\begin{itemize}
\item[1.] \underline{Single pion production in $\nu N$ scattering:}\\
In this first step a pion is produced by the scattering of the
incoming neutrino off a nucleon in the target, such that nuclear
corrections due to the Pauli principle and the Fermi motion
of the nucleons should be taken into account.
\item[2.] \underline{Multiple scattering of pions:}\\
Once the pion has been produced it will travel through the nucleus.
During this journey the pion can have several rescatterings
in which it can change its charge or be absorbed.
\end{itemize}
\vspace*{-1ex}
% Discussion of Factorization
Step 2 can be described by a $3 \times 3$ charge exchange matrix
$M$ which depends only on the properties of the target
[modeled by a charge density profile $\rho(r)$] but which is 
independent of the identities of the leptons in step 1.
It should be noted that
the above assumption of two independent steps generates
%the assumption that the two steps are independent generates
predictive power, since the formalism can be applied
to charged current (CC) and neutral current (NC) neutrino-
and electroproduction of pions in the resonance region.

% Matrix relation
Absorbing the Pauli suppression factor of step 1 into the
normalization of $M$, the measurable 
final distributions (cross sections)
of pions $\pif$ can be related to the initial distributions
$\pii$ for a {\em free} target in the simple form
\begin{equation}
{\underbrace{\left(\begin{array}{c}\displaystyle
{\der \sigma(_ZT^A;{\pi^+})\over \der Q^2\der W}\\
\displaystyle{\der \sigma(_ZT^A;{\pi^0})\over 
\der Q^2\der W}\\
\displaystyle{\der \sigma(_ZT^A;{\pi^-})\over 
\der Q^2\der W}
\end{array}\right)}_{\rm nuclear \,target}} 
%={\underbrace{\betont M}_{\rm\zitat ANP model} 
={{M}
{\underbrace{\left(\begin{array}{c}\displaystyle
{\der \sigma(N_T;{\pi^+})\over \der Q^2\der W}\\
\displaystyle{\der \sigma(N_T;{\pi^0})\over \der Q^2\der W}\\
\displaystyle{\der \sigma(N_T;{\pi^-})\over \der Q^2\der W}
\end{array}\right)}_{\rm free\, nucleon}}}
\label{eq:fac}
\end{equation}
with 
\begin{equation*}
{\der \sigma({N_T};\pm 0)\over \der Q^2\der W} 
= {{Z}{\der \sigma({p};\pm 0)
\over \der Q^2\der W}
+ {(A-Z)}{\der \sigma({n};\pm 0)
\over \der Q^2\der W}} 
%\label{eq:free}
\end{equation*}
where the free nucleon cross sections are averaged over the
Fermi momentum of the nucleons.
The matrix $M$ depends on the target material and the 
final state kinematic variables, i.e.\ $M = M[T; Q^2,W]$.
% Comment on Averaging Approximation?

% Isoscalar Target (I = 0 => N = Z)
In the following we will restrict the discussion to isoscalar 
targets and refer the reader to ref.\ \cite{Adler:1974wu} 
for targets with a neutron excess.
The charge exchange matrix 
$M$ for isoscalar targets
can be parameterized by three
independent parameters 
$A_p$, $d$, and $c$:
% $A_p[T; Q^2,W]$, $d[T; Q^2,W]$, $c[T; Q^2,W]$
\begin{equation}
{M} = { A_p} \left(\begin{array}{ccc}
{1-{ c} - { d}} & { d} & { c} \\
{ d} & {1-2 { d}} & { d} \\
{ c} & { d} & {1-{ c} - { d}}
\end{array}\right).
\label{eq:M1}
\end{equation}
So far there is no direct experimental information on these
three parameters available.
On the other hand a model for pion multiple scattering 
by Adler, Nussinov and Paschos (ANP model) 
\cite{Adler:1974qu} allows to calculate these parameters 
and (averaged) charge exchange matrices for oxygen, argon 
and iron targets can be found in \cite{Paschos:2000be}.

% Large effects 
These matrices are useful to calculate CC and NC
single pion production in the (3,3) resonance region 
induced by neutrinos 
\cite{Paschos:2000be,Paschos:2001np,Paschos:2002mb,prsy}
and electrons \cite{psy} 
and the rescattering effects are predicted to be quite large.
% Fig. 1 ???
For example, the pion energy spectra for $\pi^0$'s 
produced on oxygen targets in NC neutrino production
will be reduced by up to
$40 \%$ whereas the analogous spectra for the charged pions
remain almost unchanged compared to the free nucleon cross sections
\cite{Paschos:2000be}.
Similar results are predicted for the $W$-distributions
for 1-pion resonance production in electron-oxygen 
scattering \cite{psy}.
For the $\pi^0$ the $W$-distribution will be
reduced by about $50 \%$ while the production of $\pi^+$
remains almost unchanged compared to the free case since
the decrease due to absorption is widely compensated by an
increase due to charge exchange.
This marked signal of rescattering effects 
should be easily testable 
in low energy electron scattering experiments, for example
at JLAB.

% Outline
%
% Introduction to the ANP model
The rest of this paper is organized as follows.
In the next section we give an introduction
to the ANP model.
%
% Fixing the Normalization
In Sec.~\ref{sec:norm} we discuss the normalization
of pion absorption which we fix by experimental information.
% Dynamics of ANP model
In Sec.~\ref{sec:dynamics} 
we discuss the dynamics of the ANP model in more detail.
Finally in Sec.~\ref{sec:summary} we summarize the
main results.

\section{The ANP model} 
In this section we introduce the multiple scattering model
by Adler, Nussinov and Paschos \cite{Adler:1974qu}.
Due to space limitations the presentation will be short
and we refer the reader to Ref.\ \cite{Adler:1974qu} for
a thorough discussion of the omitted details.

The main ingredients of the model are as follows:
(i) The target nucleus is taken to be a collection of 
independent nucleons distributed spatially according to
a charge density profile $\rho(r)$.
(ii) The nucleons are regarded to be fixed within the
nucleus neglecting Fermi motion and nucleon recoil effects.
This implies that the pion energy $E_\pi$ does not change
in the elastic scatterings.
Furthermore the target is assumed to stay isotopically 
neutral ($Z=N$) during the multiple scattering processes.
(iii) The pion interactions in the nucleus are assumed to
be incoherent $\pi N$ reactions taking place in the
$\Delta$ resonance region.
Since pion production and more complex channels are closed
in this region the rescattering processes can be described by
two cross sections, the pion absorption cross section
per nucleon $\sigabs(W)$ 
and the usual elastic $\pi N$
cross sections.
The relevant $\pi N$ reactions read
\begin{gather*}
\pi^+ + N \to \pi^+ + N ,\quad \pi^+ + n \to \pi^0 + p ,
\\
\pi^0 + N \to \pi^0 + N ,\quad
\pi^0 + p \to \pi^+ + n ,
\\
\pi^0 + n \to \pi^- + p ,\quad
\pi^- + N \to \pi^- + N ,
\\
\pi^- + p \to \pi^0 + n,%\quad
\\
\pi^{\pm,0} + N \to X ,\quad (\pi \not\in X) 
{(\rightarrow {absorption})}\ .
\end{gather*}
(iv) In the $\Delta$ resonance region the $\pi N$ cross section
is dominated by the isospin $I=3/2$ amplitude such that all elastic
cross sections are related to $\sigma_{\pi^+ p}(W)$ by 
Clebsch-Gordan coefficients.
Let 
\begin{equation}
\qi = 
\big(n_i({ \pi^+}),n_i({ \pi^0}),n_i({ \pi^-})\big)^T
\end{equation}
denote the initial multiplicity distribution of pions 
in the medium (with fixed pion energy).
In a single elastic scattering this distribution will be modified
to a final distribution $\qf$:
\begin{equation}
\qf = Q\ \qi
\end{equation}
with a known $3 \times 3$ matrix $Q$ \cite{Adler:1974qu} 
which follows easily from 
a Clebsch-Gordan analysis of isospin.

Accordingly, taking into account all possible multiple scatterings
the final pion charge distribution is given by
\begin{equation}
 \qf = M\ \qi, \qquad   M = \sum_{n=0}^\infty {P_n} Q^n
\end{equation}
where $P_n$ is the  probability that the pion exits the 
nucleus after exactly $n$ 
$\pi N$ scatterings.
It should be noted that
the matrix $Q$ does not include any absorption
which is taken into account by
$\sum_{n=0}^\infty P_n < 1$.
The eigenvalues and eigenvectors of the matrix $Q$ are given by
\begin{eqnarray}
\lambda_1 &=& 1\, , \, q_1 = (1,1,1)^T, 
\\
\lambda_2 &=& \tfrac{5}{6}\, , \, q_2 = (1,0,-1)^T,
\\
\lambda_3 &=& \tfrac{1}{2}\, , \, q_3 = (1,-2,1)^T 
\end{eqnarray}
and one finds immediately the three eigenvalues of the charge exchange 
matrix $M$ in dependence
of the three eigenvalues $\lambda_k$ of $Q$
\begin{equation}
f(\lambda_k)=\sum_{n=0}^\infty P_n \lambda_k^n \, , \, k=1,2,3\ .
\label{eq:flam}
\end{equation}

The connection between the eigenbasis and the canonical basis
is given by
%\begin{eqnarray*}
%A_p\,(1-c-d) & = & \tfrac{1}{3} f(1)+\tfrac{1}{2} f(\tfrac{5}{6})
%+\tfrac{1}{6} f(\tfrac{1}{2})
%\\
%A_p\, d     & = & \tfrac{1}{3} f(1)-\tfrac{1}{3} f(\tfrac{1}{2})
%\\
%A_p\, c     & = & \tfrac{1}{3} f(1)-\tfrac{1}{2} f(\tfrac{5}{6})
%+\tfrac{1}{6} f(\tfrac{1}{2})
%\end{eqnarray*}
\begin{gather}
A_p\,(1-c-d) =  \tfrac{1}{3} f(1)+\tfrac{1}{2} f(\tfrac{5}{6})
+\tfrac{1}{6} f(\tfrac{1}{2}),
\nonumber\\
A_p\, d     =  \tfrac{1}{3} f(1)-\tfrac{1}{3} f(\tfrac{1}{2}),
\nonumber\\
A_p\, c     =  \tfrac{1}{3} f(1)-\tfrac{1}{2} f(\tfrac{5}{6})
+\tfrac{1}{6} f(\tfrac{1}{2})
\end{gather}
or inversely
\begin{eqnarray}
A_p & = &  f(1), 
\nonumber\\
c & = &  \tfrac{1}{3} - \tfrac{1}{2}{f(\tfrac{5}{6})/f(1)}
       + \tfrac{1}{6}{f(\tfrac{1}{2})/f(1)},
\nonumber\\
d &=&  \tfrac{1}{3}{\left[1 - {f(\tfrac{1}{2})/f(1)}\right] }\ .
\end{eqnarray}
As already mentioned in the introduction, the Pauli suppression
factor of step 1 will be conveniently absorbed into the normalization 
of the charge exchange matrix $M$
\begin{equation}
A_p \rightarrow A_p = g(W,Q^2) f(1)\ .
\end{equation}
The function $f(\lambda)$ contains the 
dynamical details of pion multiple scattering in the nucleus.
An outline of how this function is calculated by solving
a transport problem for pions in the nucleus 
will be given in the next section.
\subsection{Calculation of $ f(\lambda)$}
To a very good approximation
\cite{Adler:1974qu} 
the three-dimensional problem can be reduced to a one-dimensional
transport problem by projecting the forward-hemisphere
of the scattered pion onto
the forward direction and the backward-hemisphere onto the
backward direction [see Eq.~\eqref{eq:diff}].
In this approximation the pion scatters back and forth along 
its initial direction of motion until it is absorbed or leaves the
nucleus.
As is explained in more detail in Refs.~\cite{Adler:1974qu,Adler:1974wu} 
the nucleon density profile along the scattering line can 
be scaled out of the problem by an appropriate change in length
variable
such that it is equivalent to consider a {\em uniform} one-dimensional
nuclear medium extending from $x=0$ to $x=L$ with 
effective length (optical thickness)
\begin{equation}
L = L(b) = \frac{1}{\rho(0)} \int_{-\infty}^{+\infty} \der z\ 
\rho(\sqrt{z^2+b^2})\ ,
\label{eq:optical}
\end{equation}
where $b$ is the impact parameter.
Furthermore, to simplify the discussion we only consider
forward scattering of the pions (and also neglect Pauli
suppression in this step 2).
The solutions 
taking into account forward and backward scattering, $f_\pm(\lambda)$,
can be found in Appendix A of \cite{Adler:1974qu}.
Apart from being more realistic, they are important for 
estimates of the background
to proton decay by neutrino production of pions in the
$\Delta$ resonance region \cite{Gaisser:1986vn}.

The transport process can be described in terms of the 
basic probabilities
for a pion density to propagate from $y$ to 
$x$ and 
to a) interact (scattering or absorption) 
b) scatter
c) be absorbed
in $[x,x+dx]$
\begin{eqnarray}
a)\, \bra{x} \Ptot \ket{y} &=& \kappa e^{-\kappa |y-x|} \Theta(y-x)
\label{eq:pa}\\
b)\, \bra{x} \Pcex \ket{y} & =& \mu \bra{x} \Ptot \ket{y}
\label{eq:pb}\\
c)\, \bra{x} \Pabs \ket{y} & =& \abs \bra{x} \Ptot \ket{y}
\label{eq:pc}
\end{eqnarray}
with $ \kappa = \rho(0) \sigtot$ ('inverse interaction length'),
$\mu = \sigcex/\sigtot$ 
being the probability that the pion is scattered 
and
$\abs = \sigabs/\sigtot$ the  
probability that the pion is absorbed (in a single scattering process).
The choice of $\rho(0)$ as density of the uniform one-dimensional
medium (i.e. $\kappa = \rho(0) \sigtot$)
is related to the
normalization of the effective length in Eq.~\eqref{eq:optical}.
Note also that $\mu + \abs = 1$ since $\sigtot = \sigcex + \sigabs$.
The cross sections for charge exchange (cex) and 
absorption (abs) will be specified below.

% calculation of dynamical function $ f(\lambda)$:
The density of pions in the medium after 
$n$ scatterings, $\ket{\rin{n}}$,
is related to the initial density $\ket{\rin{0}}$ by
\begin{equation}
\ket{\rin{n}} = \Pcex^n \ket{\rin{0}}
\end{equation}
where the initial density is normalized 
to $\braket{x}{\rin{0}} = 1/L$
such that the number
of pions in the medium is one:
\begin{equation}
\Nin{0} \equiv \int_0^L \der x\ \braket{x}{\rin{0}} = 1 \ .
\end{equation}
The density of pions leaving the medium after $n$ 
scatterings, $ \ket{\rout{n}}$,
is given by 
\begin{eqnarray}
\ket{\rout{n}} & =& (1-\Ptot) \ket{\rin{n}}
\\
& =& (1-\Ptot) \Pcex^n \ket{\rin{0}}\ .
\end{eqnarray}
The probability that the pion leaves the medium after
exactly $n$ rescatterings is then
\begin{equation}
P_n \equiv \Nout{n} = \int_0^L\ \der x\ \braket{x}{\rout{n}}\ .
\label{eq:pn}
\end{equation}
The dynamical function $f(\lambda)$ we wish to calculate 
is then formally given by
\begin{equation}
f(\lambda) =\sum_{n=0}^\infty P_n \lambda^n 
= \int_0^L\ \der x\ \braket{x}{\psiout}
\label{eq:f}
\end{equation}
with
\begin{equation}
\ket{\psiout} := \sum_{n=0}^\infty \lambda^n \ket{\rout{n}}\ .
\end{equation}
Using $1-\Ptot = 1-\tfrac{1}{\sigma} + \tfrac{1}{\sigma} (1 -\lambda \Pcex)$
with $\sigma := \lambda \mu$
and $\sum_{n=0}^\infty \lambda^n \Pcex^n = (1- \lambda \Pcex)^{-1} =: 1+F$
we can write $\ket{\psiout}$ as
\begin{equation}
\ket{\psiout} = \left[1 + (1-\tfrac{1}{\sigma}) F\right]\ket{\rin{0}}\ .
\label{eq:out}
\end{equation}
Inserting Eq.\ \eqref{eq:out} into Eq.\ \eqref{eq:f} we find
\begin{equation}
f(\lambda) = 1 + (1-\tfrac{1}{\sigma})
\int_0^L\ \der x\ \der y\ \bra{x}F\ket{y} \frac{1}{L}
\label{eq:f2}
\end{equation}
where we have inserted the unit operator
\begin{equation}
1 = \int_0^L \der y\ \ket{y}\bra{y}
\end{equation}
and used $\braket{x}{\rin{0}} = 1/L$.

Using the definition of the operator $1+F=(1-\lambda \Pcex)^{-1}$ and 
$(1-\lambda \Pcex)^{-1} (1 -\lambda \Pcex)=1$ we can furthermore
write
\begin{equation}
F = \lambda \Pcex + F\ \lambda \Pcex
\end{equation}
which leads us, using Eqs.\ \eqref{eq:pa} and \eqref{eq:pb},
to the transport integral equation
\begin{equation}
f(y) = \sigma (1 - e^{-\kappa y})
+ \sigma \kappa \int_0^y \der z\ f(z) e^{-\kappa (y-z)}
\label{eq:transport}
\end{equation}
with 
\begin{equation}
f(y) := \int_0^L \der x\ \bra{x} F \ket{y}\ .
\label{eq:fy}
\end{equation}
The integral equation can be transformed into a differential equation
by evaluating $\tfrac{\der}{\der y} [e^{\kappa y} f(y)]$ for the 
left and the right side of Eq.\ \eqref{eq:transport} resulting
in
\begin{equation}
\frac{\der}{\der y}f = \kappa (\sigma - 1) f(y) + \kappa \sigma
\, , \, f(0) = 0\ .
\label{eq:dgl}
\end{equation}
This differential equation is solved by
\begin{equation}
f(y) = \frac{\sigma}{1-\sigma} \left[1-h(y)/h(0) \right]\ ,
\label{eq:solution}
\end{equation}
where $h(y) = e^{\kappa (\sigma - 1)y}$ is a solution of the
homogeneous equation $h^\prime = \kappa (\sigma - 1) h$.
The final solution for the dynamical function $f(\lambda)$
is now easily found to be
\begin{eqnarray}
f(\lambda) &=& 1 + (1-\tfrac{1}{\sigma})
\frac{1}{L} \int_0^L\ \der y\ f(y)
\nonumber\\
&=& \frac{1 - e^{-\kappa L(1-\sigma)}}{\kappa L (1-\sigma)}\, , \, 
\sigma = \lambda \mu\ .
\label{eq:final}
\end{eqnarray}
Finally, we average the solution in Eq.\ \eqref{eq:final} 
over impact parameters: 
\begin{equation}
f(\lambda) = 
\frac{\int_0^{\infty} b\ \der b\ L(b) 
f(\lambda,L(b))}{\int_0^{\infty} b\ \der b\ L(b)}\ .
\label{eq:impactav}
\end{equation}

%\subsection{ANP model: final results}
%{
%\begin{equation*}
%M_{\pm} = { A_{\pm}}  \left(\begin{array}{ccc}
%{1-{ c_{\pm}} - { d_{\pm}}} & { d_{\pm}} & 
%{ c_{\pm}} \\
%{ d_{\pm}} & {1-2 { d_{\pm}}} & { d_{\pm}} \\
%{ c_{\pm}} & { d_{\pm}} & {1-{ c_{\pm}} - 
%{ d_{\pm}}}
%\end{array}\right)
%\end{equation*}
%$ { A_{\pm}} = g(W,Q^2) { f_{\pm}(1)}$,\\
%$ { c_\pm} = \tfrac{1}{3} 
%- \tfrac{1}{2} { f_\pm(\tfrac{5}{6})/f_\pm(1)}
%+ \tfrac{1}{6} { f_\pm(\tfrac{1}{2})/f_\pm(1)}
%$,\\
%$ { d_\pm} = \tfrac{1}{3}[1 
%- { f_\pm(\tfrac{1}{2})/f_\pm(1)}]$\\
%with
%\begin{equation*}
%f_{\pm}(\lambda) = 
%\frac{\int_0^{\infty} b\ db\ L(b) 
%f_{\pm}(\lambda,L(b))}{\int_0^{\infty} b\ db\ L(b)}
%\end{equation*}
%}
%
%\underline{Final result for the dynamical functions $ f(\lambda,L)$:}
%\begin{eqnarray*}
%f & =& 
%{\frac{e^{\kappa \sigma L} - 1}{\kappa \sigma L}}
%\frac{1+ \mu e^{-\kappa \sigma L}}{1+\mu}
%\\
%& =& f_{+}+ f_{-}
%\\
%f_{+} & =& 
%{ \frac{e^{\kappa \sigma L} - 1}{\kappa \sigma L}}
%\frac{{ \mu^2} e^{-\kappa \sigma L}-1}{\mu^2-1}
%\\
%f_{-} & =& 
%{ \frac{e^{\kappa \sigma L} - 1}{\kappa \sigma L}}
%\frac{{ \mu}(1- e^{-\kappa \sigma L})}{\mu^2-1}
%\end{eqnarray*}
%with 
%$\sigma = \sqrt{(1-\sigma_+)^2 - \sigma_{-}^2}$, 
%$\mu = \tfrac{\sigma+1-\sigma_+}{\sigma_-}e^{\kappa \sigma L}$, and
%$\sigma_\pm = { \lambda} \mu_\pm$.
%
%Limiting case: only forward scattering ($\mu_{-} = 0$)
%\begin{equation*}
%f_{-} = 0, \qquad f_{+} =
%\frac{1-e^{-\kappa L(1-\lambda \mu_+)}}{\kappa L (1-\lambda \mu_+)}
%\end{equation*}

%
\subsection{ANP model: Input}
In this subsection we summarize the input quantities
entering the ANP model.

\begin{figure}
\epsfig{file= 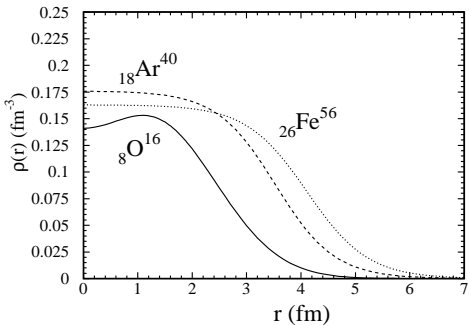,height=5.cm,width=0.9\linewidth}
\vspace*{-1cm}
\caption{Charge density profiles for oxygen, argon, and iron
normalized to $\int \der^3 r \rho(r) = A$.}
\label{fig:rho}
\end{figure}

(i) First of all the nucleon density $\rho(r)$ enters
the calculation of the effective length $L(b)$ via
$ L(b) = \tfrac{1}{\rho(0)} \int \der z \ \rho[r(z,b)]$.
The charge density profiles have been determined in electron
scattering experiments.
For lighter targets like $_{8}O^{16}$ 
the charge density is 
parameterized by a 'harmonic oscillator form' 
\begin{equation}
\rho(r) = \rho(0)\exp(-r^2/R^2) (1+C \frac{r^2}{R^2} +
C_1 \frac{r^4}{R^4}) 
\end{equation}
whereas for heavier targets like $_{18}Ar^{40}$ and $_{26}Fe^{56}$ 
a 'two parameters Fermi model' is utilized:
\begin{equation}
\rho(r) = \rho(0)\big[1+\exp{((r-C)/C_1)}\big]^{-1} \ .
\end{equation}
The parameters $\rho(0)$, $R$, $C$, and $C_1$ can be found 
in Table 1 of \cite{Paschos:2000be}.
The corresponding charge density profiles $\rho(r)$
are shown in Fig.~\ref{fig:rho}.
%{Atom. Data Nucl. Data Tabl. {\bf 36} ('87) 495}
%and are listed in Table~\ref{tab:1}.
%\begin{table*}[ht]
%\caption{Parameters for charge density profiles.}
%\label{tab:1}
%\renewcommand{\tabcolsep}{0.3pc} % enlarge column spacing
%\renewcommand{\arraystretch}{0.8}  % enlarge line spacing
%%\begin{center}
%\scriptsize
%\begin{tabular}{cccccc}
%$_ZT^A$ & $a[fm]$ & $C[fm]$ & $C_1[fm]$ & $R[fm]$ & $\rho(0)[fm^{-3}]$ \\ 
%\hline
%$_{8}O^{16}$ & 2.718 & 1.544 & 0 & 1.833 & 0.141\\
%$_{18}Ar^{40}$ & 3.393 & 3.530 & 0.542 & 4.380 & 0.176\\
%$_{26}Fe^{56}$ & 3.801 & 4.111 & 0.558 & 4.907 & 0.163\\
%\end{tabular} 
%%\end{center}
%%\\
%%$a = \sqrt{<r^2>}$ (root-mean-square radius)
%%2PFM: $R^2 = \tfrac{5}{3} <r^2>$
%%(Normalization: $\int d^3 x\ \rho(r) = A$)
%\end{table*}
%
%

% elastic $\pi N$ scattering in the $(3,3)$ region:
(ii) The second input is the usual cross section for
elastic $\pi N$ scattering in the $(3,3)$ resonance region.
This region is dominated by the $I = \tfrac{3}{2}$ amplitude.
The elastic cross sections per nucleon for a $\pi^k$ ($k=\pm,0$)
in the initial state
are given by $Z/A\ \sigma_{\pi^k p} + N/A\ \sigma_{\pi^k n}$.
For an isoscalar target ($Z=N$) and neglecting 
the $I=\tfrac{1}{2}$ channel 
they are independent of the pion charge
and will be denoted by $\sigcex$.
Moreover,
the elastic cross section $\sigcex$ is proportional
to the cross section $\sigma_{\pi^+ p}(W)$
\begin{eqnarray}
\frac{\der \sigma}{\der \Omega} &\propto& \sigma_{\pi^+ p}(W)\ 
(1+3 \cos^2 \phi)\ ,
%\ \underbrace{h(W,\phi)}_{\rm Pauli\ fac.}
\label{eq:diff}\\
\sigcex &=& 
\tfrac{2}{3} \sigma_{\pi^+ p}(W)
%\ [h_{+}(W)+h_{-}(W)]
\label{eq:sigcex}
\end{eqnarray}
with
\begin{equation}
\sigma_{\pi^+ p}(W) = \sigma_{(3,3)}(W)  + 20\ {\rm mb}
\end{equation}
where $\sigma_{(3,3)}(W)$ is a resonant cross section
which can be found in \cite{Adler:1974qu} and
the second term is a constant non-resonant background.
In Eqs.~\eqref{eq:diff} and \eqref{eq:sigcex} we have omitted
for simplicity 
a possible Pauli suppression factor taking into
account the Pauli exclusion principle in step 2.
Note also, that it is 
debatable
whether such a factor 
should be included here
since we are working in a picture with fixed nucleons.

\begin{figure}[th]
%\begin{minipage}[b]{.8\linewidth}
\epsfig{file=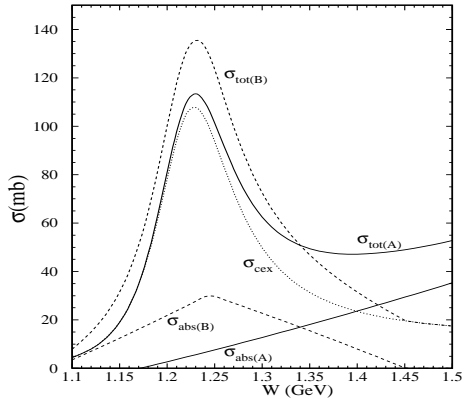,height=6cm,width =0.9\linewidth }
%\end{minipage}
\vspace*{-1cm}
\caption{Total, elastic, and absorption cross sections in
the $(3,3)$ region.}
\label{fig:xs}
\end{figure}

% absorption:
(iii) The third ingredient is the cross section per nucleon
for pion absorption, $\sigabs(W)$.
For this quantity Sternheim and Silbar have published
two parametrizations \cite{Sternheim:1972ad,Silbar:1973em}
which they have
extracted from data on single pion production in $p A$ scattering.
These parametrizations can also be found in \cite{Adler:1974qu}.
In Fig.\ \ref{fig:xs} these two absorption cross sections,
model (A) and model (B), are shown together with the elastic
and the corresponding total cross sections.
As can be seen the
models (A) and (B) for $ \sigabs$ are quite different in {\em shape} 
and {\em normalization}.

%\underline{dependent quantities:}
%\begin{itemize}
%\item $ \sigtot(W) = \sigabs(W) + \sigcex(W)$
%\item $ \kappa = \rho(0) \sigtot$: 
%{inverse free path length}
%\item $ \mu_{\pm} = \tfrac{1}{3} \sigma_{\pi^+ p}(W)\ 
%h_{\pm}(W)/\sigtot$:
%\\ 
%{probability for charge exchange} in a {\em single} 
%$\pi N$ scattering
%\item $ a = \sigabs(W)/\sigtot(W)$:
%\\ 
%{probability for absorption} in a {\em single} 
%$\pi N$ scattering
%\item $\mup + \mum + a = 1$
%\end{itemize}

\section{Fixing the normalization of $\sigabs(W)$}
\label{sec:norm}
The two models for the absorption cross section,
model (A) \cite{Sternheim:1972ad} and model (B) \cite{Silbar:1973em}
result in quite different total amounts of absorbed pions:
for example,
model (A) predicts that about $19\%$ 
of the pions will be absorbed in oxygen
compared to $43\%$ in model (B).

These numbers have been obtained in an 'averaging approximation'
\cite{Adler:1974qu}
in which the $W$-dependence of $f(\lambda, W)$ is 
averaged over the $(3,3)$ region.
%\begin{equation}
% \fbar(\lambda) = \frac{\int \der W\ q^{-1}(W) 
%\sigma_{(3,3)}(W)\ {f(\lambda,W)}}
%{\int \der W\ q^{-1}(W) \sigma_{(3,3)}(W)}\ .
%\end{equation}
In this case 
the dynamical functions (the charge exchange matrix) 
are
mainly sensitive to the region around $W \simeq m_\Delta$
and thus mainly sensitive to the {\em normalization} of 
$\sigabs(W \simeq m_\Delta)$.

In the following we fix this normalization by using
experimental data for 1-pion production in CC
$\nu_\mu$-deuteron [$\leftrightarrow$ free case] 
and $\nu_\mu$-neon scattering.
These data have been
weighted to the {\em same} atmospheric $E_\nu$ 
spectrum in a paper by Merenyi et al.\ \cite{Merenyi:1992gf}
allowing for direct comparison.
In Table 1 of \cite{Merenyi:1992gf} relative charged current 
populations (w.r.t.~the total cross section) are provided
for deuteron (D) and neon (Ne).
The fractional contributions of the 1-pion production channels
$(\pi^+,\pi^0,\pi^-)$ are given by:
\begin{eqnarray}
\text{D}:\, \ri &=& (0.165,0.09,0)^T 
\\
\text{Ne}:\, \rf 
&=& (0.11,0.05,0.01)^T 
\nonumber\\
& & \pm (0.014,0.02,0.01)^T\, .
\end{eqnarray}

The fractional populations $\ri$ and $\rf$
are related in the ANP model by:
\begin{equation}
\rf = M \ri \times K\, ,\quad
K = \frac{10 \sigma_{\rm tot}(\text{D})}{\sigma_{\rm tot}(\text{Ne})}\ .
\end{equation}
Unfortunately the total cross sections, weighted to an atmospheric
neutrino flux, have not been published in \cite{Merenyi:1992gf}.
However, the correction factor $K$ should be close to one
since the total cross section for Ne is not affected
by charge exchange and pion absorption.
In the following exercise we assume $K=1$ and later assign
a normalization uncertainty of $3\%$.

Furthermore, it is not viable to solve 
$\rf \overset{!}{=} {M[A_p,d,c]}\ \ri$
for the three parameters $ A_p,d,c$ since the solution
{\em strongly} varies within the errors of $\rf$.
Instead, we have made the reasonable assumption $0 \le c < d$
which means that the
probability for $\pi^- \to \pi^+$ is smaller than the probability for 
$\pi^0 \to \pi^+$.
Under this assumption one can fit the two parameters $A_p$ and
$d$ for a fixed value of $c$ as long as $c < d$:
%\begin{table*}[ht]
%\caption{Parameters for charge density profiles.}
%\label{tab:2}
%\renewcommand{\tabcolsep}{0.3pc} % enlarge column spacing
%\renewcommand{\arraystretch}{0.8}  % enlarge line spacing
%\begin{center}
%\scriptsize
\begin{tabular}{cccc}
$c$ & & $A_p$ & $d$ \\
\hline
$c=0\phantom{.00}$ & $\rightarrow$ Fit: & 0.695 & 0.147\\
$c=0.01$ & $\rightarrow$ Fit: & 0.696 & 0.128\\
$c=0.02$ & $\rightarrow$ Fit: & 0.696 & 0.109\\
$c=0.03$ & $\rightarrow$ Fit: & 0.696 & 0.091\\
$c=0.04$ & $\rightarrow$ Fit: & 0.697 & 0.072\\
$c=0.05$ & $\rightarrow$ Fit: & 0.698 & 0.053\\
\end{tabular} 
%\end{table*}
\newline
\noindent [$ \chi^2/d.o.f \simeq 0.4$]
%
% Observation:
\newline
As can be seen the 
parameter $ A_p$ is well constrained:
$ A_p = 0.696 \pm 0.002$.
More conservatively we use in the following
$A_p = 0.70 \pm 0.02$ taking into account the normalization uncertainty
due to the correction factor $K$.
On the other hand
the parameters $d$ and $c$ are correlated and adopt values in the
range
$c \in [0,0.05]$, $d \in [0.15,0.05]$.

% Fraction $A$ of absorbed pions:
Inspecting Eq.\ \eqref{eq:flam} we find
that $f(\lambda = 1) = \sum_{n=0}^\infty P_n = 1-A$ where
$A$ is the fraction of absorbed pions.  
Therefore, 
the averaged fraction $\bar A$ of absorbed pions 
is given by
$\bar A = 1-\fbar(\lambda=1)$
where $\fbar(\lambda)$ is
the averaged dynamical function which is related
to the above extracted parameter $A_p$ via
$A_p = g(\bar W,\bar Q^2) \fbar(1)$.
In order to estimate the effect of the 
Pauli suppression factor
we take $ \bar W \simeq m_\Delta$ and
$\bar Q^2 \simeq 0.1\ \gevsq$ leading to
$g(\bar W,\bar Q^2) = 0.93 \pm 0.05$
[see Table 3 in \cite{Paschos:2000be}].
Using these values we finally find $\fbar(1) = 0.75 \pm 0.05$
implying
that about $25\%$ of the pions have been absorbed in the neon target:
$\bar A = 0.25 \pm 0.05$.

% Fixing the normalization of $ \sigabs(W)$
A fraction of $ 25 \%$ pion absorption can be 
obtained by renormalizing absorption
model (B) [model (A)] by a factor $\simeq 0.3$ 
[$\simeq 1.4$].
With this renormalization we can compare the ANP model
with the above fitted results.
The ANP model gives for both renormalized absorption
models for a neon target
$\fbar(1) =0.75$ (by construction), $d = 0.15$,
and $c = 0.05$.
This result compares favorably with the fitted values.

\section{Dynamics of the ANP model}
\label{sec:dynamics}
\subsection{Linearisation}
\label{sec:linear}
The simple solution in Eq.\ \eqref{eq:final}
is well-suited for further analytical investigations
of pion multiple scattering.
For example, the forward solution \eqref{eq:final}
at $\lambda = 1$,
\begin{equation}
f(\lambda=1,L,W)= 
\frac{1-e^{-\rho_0 L \sigabs}}{\rho_0 L \sigabs}\ ,
\end{equation}
can be considered 
in the limit $ \rho_0 L \sigabs \ll 1$
relating the fraction of absorbed pions $A$
to the absorption cross section per nucleon $ \sigabs(W)$:
\begin{equation}
A(L,W)= 1 - f(1,L,W) \simeq \tfrac{1}{2} \rho_0 L \sigabs(W)\ .
\label{eq:abs1}
\end{equation}

% For Oxygen
Averaging $L=L(b)$ over the impact parameters $b$ 
for oxygen as target material
gives 
$\bar L \simeq 1.9 R$ with radius $ R\simeq 1.833\ {\rm fm}$. 
The nuclear density for oxygen,
$ \rho_0 = 0.141\ {\rm fm}^{-3}$,
has been taken from Table 1 in \cite{Paschos:2000be}. %\ref{tab:1}.
Replacing $\rho_0 L$ by $\rho_0 \bar L \simeq 0.05\ {\rm mb}^{-1}$
in Eq.~\eqref{eq:abs1} 
% has to be weighted by the number of pions at a given W; averaging approx.!
and taking into account the averaging over the
$(3,3)$ resonance by the replacement $W \to m_\Delta$
we arrive at the following 
rule of thumb for the fraction of absorbed pions in oxygen
\begin{equation}
A \simeq 0.025\ \sigabs(W \simeq m_\Delta) [{\rm mb}]\ .
\label{eq:abs2}
\end{equation}

The absorption cross sections can be taken from Fig.\ \ref{fig:xs}
and we find the following fractions of absorbed pions:\newline
%\begin{table*}[ht]
%\caption{Parameters for charge density profiles.}
%\label{tab:2}
%\renewcommand{\tabcolsep}{0.3pc} % enlarge column spacing
%\renewcommand{\arraystretch}{0.8}  % enlarge line spacing
%\begin{center}
%\scriptsize
\begin{tabular}{ccc}
model &$\sigabs(W=m_\Delta)$& $A$\\
\hline
(A)& $\simeq 6.0\ {\rm mb}$ & $15\%$\\
(B)& $\simeq 28.4\ {\rm mb}$ & $71\%$\\
0.3 $\times$ (B) & $\simeq 8.5\ {\rm mb}$ & $21\%$\\
\end{tabular} 
\newline
Here the renormalization factor $0.3$ for model (B) has
been taken from the preceding section.
The same result is found by renormalizing model (A) with
a factor $1.4$.
Note also that the $71\%$ for (the original) model (B)
is not realistic because the condition 
$\rho_0 \bar L \sigabs \ll 1$ is not satisfied.
\subsection{'How many' multiple scatterings?}
\label{sec:prob}

\begin{figure}
%\vspace*{-2cm}
%\begin{minipage}[b]{.8\linewidth}
\epsfig{file=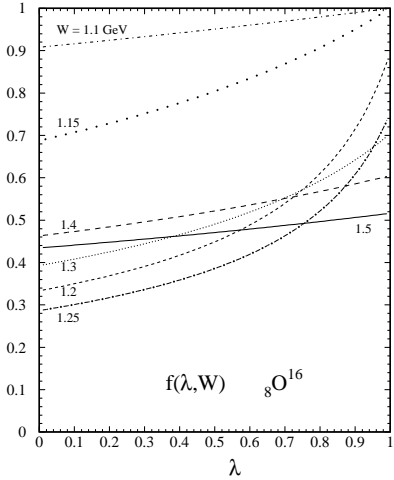,height = 7cm}
%\end{minipage}
\vspace*{-1cm}
\caption{Dynamical function $f(\lambda,W)$ in dependence of $\lambda$
for several values of $W$.}
\label{fig:3}
\end{figure}

Figure \ref{fig:3} shows 
the dynamical function $f(\lambda,W)$ for oxygen 
in dependence of $\lambda$ for several values of $W$.
Writing $f(\lambda) = \sum_{n=0}^\infty P_n \lambda^n$
with the probabilities $P_k$ ($k=0,1,2,\ldots$) that
the pion is observed after $k$ $\pi N$ scatterings
it is obvious that
$ f(\lambda=0) = P_0$ and $ f(\lambda=1)= 1-A$
where $A$ is the probability for pion absorption.
[Note that $\sum_{k=0}^\infty P_k + A = 1$.]
It is an interesting question, how large the
probabilities $P_n, n>0$ for multiple scattering are.
Qualitatively, 
a stronger curvature of the function $f(\lambda)$
indicates a higher probability for multiple scattering.
As can be seen, in the vicinity of $W=m_\Delta$ the
probability that the pion rescatters several time is
largest whereas at $W=1.1\ \gev$ the curve is almost linear
such that only $P_0$ and $P_1$ contribute appreciably.

\begin{table}
{%\scriptsize
\renewcommand{\arraystretch}{0.8}  
\caption{Probabilities in $\%$ for multiple scattering.}
\begin{tabular}{p{1.4cm}|c|c|c|c}
$W [\gev]$ & 1.1 & 1.2  & 1.25 & 1.5\\
\hline\hline
$A$        & {0.0} & 11.8 & {25.7} & 48.4 \\ 
$P_0$      &{90.8} & 33.6 & 28.8          & 43.4 \\
$P_1$      & 8.0          & 14.1 & 12.2          &  6.5 \\
\hline
$P_2$      & 1.0          & 10.5 &  8.9          &  1.3  \\
$P_3$      & 0.13         &  7.8 &  6.5          &  0.26 \\
$P_4$      & $\ldots$     &  5.8 &  4.8          &  $\ldots$     \\
$P_5$      &              &  4.3 &  3.5          &       \\
$P_6$      &              &  3.2 &  2.6          &       \\
$P_7$      &              &  2.4 &  1.9          &       \\
$P_8$      &              &  1.8 &  1.4          &       \\
$P_9$      &              &  1.3 &  1.0          &       \\
$P_{10}$   &              &  1.0 &  0.7          &       \\
\end{tabular}
}
%\vspace*{-0.5cm}
\label{tab:prob}
\end{table}

Of course the probabilities $P_n$ can be calculated exactly
within the ANP model according to Eq.\ \eqref{eq:pn}
or by differentiating the solution for $f(\lambda)$ in Eq.~\eqref{eq:final}:
$P_n = \tfrac{1}{n!} \tfrac{\der^n}{\der \lambda^n} f_{|\lambda=0}$.

On the other hand, inspired by the fact that the pion energy
remains constant, it is interesting to make
the assumption (which is only asymptotically correct)
that $P_{k+1} \simeq r P_k$ for $k \ge 1$ or equivalently
$P_k \simeq P_1 r^{k-1}$ for $k \ge 1$. 
Under this assumption $f(\lambda)$ is a geometrical
series and is given by the simple solution
\begin{equation}
f(\lambda) = \sum_{n=0}^\infty P_n \lambda^n \simeq
P_0 + \frac{P_1 \lambda}{1 - r \lambda}\ .
\label{eq:series}
\end{equation}
The constant $r$ can be fixed from $f(\lambda=1) = 1 - A$
resulting in $r =1-\frac{P_1}{1-A-P_0}$.

Thus, once $A$, $P_0$ and $P_1$ are known, all higher 
probabilities and the function $f(\lambda)$
can be easily estimated.
The result of such a procedure is listed in the following table
where $P_0$, $P_1$ and $A$ have been determined from Fig.~\ref{fig:3}.

\section{Conclusions}
\label{sec:summary}
In this article, we have given a basic introduction to the
pion multiple scattering model of Adler, Nussinov, and Paschos
\cite{Adler:1974qu} focusing on the input parameters of the model, 
particularly on the cross section for pion absorption $\sigabs(W)$
which is the least well determined ingredient.
Using data for $\nu_\mu$-deuteron and $\nu_\mu$-neon scattering
\cite{Merenyi:1992gf} we could fix the normalization of $\sigabs(W)$
corresponding to a fraction of $(25 \pm 5)\%$ of absorbed pions in neon.
The parameters $A_p$, $d$, and $c$ have been determined to be
$A_p = 0.70 \pm 0.02$, $c \in [0,0.05]$, and $d \in [0.15,0.05]$
which compares quite favorably with the predictions of the ANP model
for neon.

% Outlook
In order to test and improve the ANP model it will
be necessary to make detailed measurements of single pion
electroproduction in the region of the $(3,3)$ resonance using
different heavy targets and to compare it with the corresponding
cross sections on free nucleons \cite{psy}.


\begin{thebibliography}{10}
\expandafter\ifx\csname bibnamefont\endcsname\relax
  \def\bibnamefont#1{#1}\fi
\expandafter\ifx\csname bibfnamefont\endcsname\relax
  \def\bibfnamefont#1{#1}\fi
\expandafter\ifx\csname url\endcsname\relax
  \def\url#1{\texttt{#1}}\fi
\expandafter\ifx\csname urlprefix\endcsname\relax\def\urlprefix{URL }\fi
\expandafter\ifx\csname bibinfo\endcsname\relax \def\bibinfo#1#2{#2}\fi
\expandafter\ifx\csname eprint\endcsname\relax \def\eprint#1{#1}\fi

\bibitem{Adler:1974qu}
\bibinfo{author}{\bibfnamefont{S.~L.} \bibnamefont{Adler}},
  \bibinfo{author}{\bibfnamefont{S.}~\bibnamefont{Nussinov}}, \bibnamefont{and}
  \bibinfo{author}{\bibfnamefont{E.~A.} \bibnamefont{Paschos}},
  \bibinfo{journal}{Phys. Rev.} \textbf{\bibinfo{volume}{D9}},
  \bibinfo{pages}{2125} (\bibinfo{year}{1974}).

\bibitem{Adler:1974wu}
\bibinfo{author}{\bibfnamefont{S.~L.} \bibnamefont{Adler}},
  \bibinfo{journal}{Phys. Rev.} \textbf{\bibinfo{volume}{D9}},
  \bibinfo{pages}{2144} (\bibinfo{year}{1974}).

\bibitem{Paschos:2000be}
\bibinfo{author}{\bibfnamefont{E.~A.} \bibnamefont{Paschos}},
  \bibinfo{author}{\bibfnamefont{L.}~\bibnamefont{Pasquali}}, \bibnamefont{and}
  \bibinfo{author}{\bibfnamefont{J.~Y.} \bibnamefont{Yu}},
  \bibinfo{journal}{Nucl. Phys.} \textbf{\bibinfo{volume}{B588}},
  \bibinfo{pages}{263} (\bibinfo{year}{2000}).

\bibitem{Paschos:2001np}
\bibinfo{author}{\bibfnamefont{E.~A.} \bibnamefont{Paschos}} \bibnamefont{and}
  \bibinfo{author}{\bibfnamefont{J.~Y.} \bibnamefont{Yu}},
  \bibinfo{journal}{Phys. Rev.} \textbf{\bibinfo{volume}{D65}},
  \bibinfo{pages}{033002} (\bibinfo{year}{2002}).

\bibitem{Paschos:2002mb}
\bibinfo{author}{\bibfnamefont{E.~A.} \bibnamefont{Paschos}},
  \bibinfo{journal}{Nucl. Phys. Proc. Suppl.} \textbf{\bibinfo{volume}{112}},
  \bibinfo{pages}{89} (\bibinfo{year}{2002}).

\bibitem{prsy}
\bibinfo{author}{\bibfnamefont{E.~A.} \bibnamefont{Paschos}},
  \bibinfo{author}{\bibfnamefont{D.~P.} \bibnamefont{Roy}},
  \bibinfo{author}{\bibfnamefont{I.}~\bibnamefont{Schienbein}},
  \bibnamefont{and} \bibinfo{author}{\bibfnamefont{J.-Y.} \bibnamefont{Yu}},
  \emph{\bibinfo{title}{Muon Spectra of Quasi-Elastic and 1-Pion Production
  Events in LBL Neutrino Oscillation Experiments}}, \eprint{hep-ph/0307223}.

\bibitem{psy}
\bibinfo{author}{\bibfnamefont{E.~A.} \bibnamefont{Paschos}},
  \bibinfo{author}{\bibfnamefont{I.}~\bibnamefont{Schienbein}},
  \bibnamefont{and} \bibinfo{author}{\bibfnamefont{J.~Y.} \bibnamefont{Yu}},
  \bibinfo{note}{{in preparation}}.

\bibitem{Gaisser:1986vn}
\bibinfo{author}{\bibfnamefont{T.~K.} \bibnamefont{Gaisser}},
  \bibinfo{author}{\bibfnamefont{M.}~\bibnamefont{Nowakowski}},
  \bibnamefont{and} \bibinfo{author}{\bibfnamefont{E.~A.}
  \bibnamefont{Paschos}}, \bibinfo{journal}{Phys. Rev.}
  \textbf{\bibinfo{volume}{D33}}, \bibinfo{pages}{1233} (\bibinfo{year}{1986}).

\bibitem{Sternheim:1972ad}
\bibinfo{author}{\bibfnamefont{M.~M.} \bibnamefont{Sternheim}}
  \bibnamefont{and} \bibinfo{author}{\bibfnamefont{R.~R.}
  \bibnamefont{Silbar}}, \bibinfo{journal}{Phys. Rev.}
  \textbf{\bibinfo{volume}{D6}}, \bibinfo{pages}{3117} (\bibinfo{year}{1972}).

\bibitem{Silbar:1973em}
\bibinfo{author}{\bibfnamefont{R.~R.} \bibnamefont{Silbar}} \bibnamefont{and}
  \bibinfo{author}{\bibfnamefont{M.~M.} \bibnamefont{Sternheim}},
  \bibinfo{journal}{Phys. Rev.} \textbf{\bibinfo{volume}{C8}},
  \bibinfo{pages}{492} (\bibinfo{year}{1973}).

\bibitem{Merenyi:1992gf}
\bibinfo{author}{\bibfnamefont{R.}~\bibnamefont{Merenyi}} \emph{et~al.},
  \bibinfo{journal}{Phys. Rev.} \textbf{\bibinfo{volume}{D45}},
  \bibinfo{pages}{743} (\bibinfo{year}{1992}).

\end{thebibliography}
\end{document}